\documentclass[acmlarge, nonacm]{acmart}

\AtBeginDocument{%
  }

\usepackage{listings}
\usepackage{color}
\usepackage{pifont}
\usepackage{algorithmic}
\usepackage{multicol}
\usepackage{multirow}

\definecolor{mygreen}{rgb}{0,0.6,0}
\definecolor{mygray}{rgb}{0.5,0.5,0.5}
\definecolor{mymauve}{rgb}{0.58,0,0.82}

\begin{document}

\title{Asynchronous Memory Access Unit: Exploiting Massive Parallelism for Far Memory Access}

\titlenote{The manuscript has been submitted to ACM TACO (Transactions on Architecture and Code Optimization) and has been accepted. The final version will be published in TACO.}

\author{Luming Wang}
\email{wangluming@ict.ac.cn}
\affiliation{%
  \institution{Institute of Computing Technology, Chinese Academy of Sciences}
  \city{Beijing}
  \country{China}}
\affiliation{%
  \institution{School of Computer Science and Technology, University of Chinese Academy of Sciencess}
  \city{Beijing}
  \country{China}}

\author{Xu Zhang}
\email{zhangxu19s@ict.ac.cn}
\affiliation{%
  \institution{Institute of Computing Technology, Chinese Academy of Sciences}
  \city{Beijing}
  \country{China}}
\affiliation{%
  \institution{School of Computer Science and Technology, University of Chinese Academy of Sciencess}
  \city{Beijing}
  \country{China}}

\author{Songyue Wang}
\email{wangsongyue22s@ict.ac.cn}
\affiliation{%
  \institution{Institute of Computing Technology, Chinese Academy of Sciences}
  \city{Beijing}
  \country{China}}
\affiliation{%
  \institution{School of Computer Science and Technology, University of Chinese Academy of Sciencess}
  \city{Beijing}
  \country{China}}

\author{Zhuolun Jiang}
\email{jiangzhuolun22z@ict.ac.cn}
\affiliation{%
  \institution{Institute of Computing Technology, Chinese Academy of Sciences}
  \city{Beijing}
  \country{China}}
\affiliation{%
  \institution{School of Computer Science and Technology, University of Chinese Academy of Sciencess}
  \city{Beijing}
  \country{China}}

\author{Tianyue Lu}
\email{lutianyue@ict.ac.cn}
\affiliation{%
  \institution{Institute of Computing Technology, Chinese Academy of Sciences}
  \city{Beijing}
  \country{China}}
\affiliation{%
  \institution{School of Computer Science and Technology, University of Chinese Academy of Sciencess}
  \city{Beijing}
  \country{China}}

\author{Mingyu Chen}
\email{cmy@ict.ac.cn}
\affiliation{%
  \institution{Institute of Computing Technology, Chinese Academy of Sciences}
  \city{Beijing}
  \country{China}}
\affiliation{%
  \institution{School of Computer Science and Technology, University of Chinese Academy of Sciencess}
  \city{Beijing}
  \country{China}}
  
\author{Siwei Luo}
\email{luosiwei@huawei.com}
\affiliation{%
  \institution{Huawei}
  \country{China}}
 
\author{Keji Huang}
\email{huangkeji@huawei.com}
\affiliation{%
  \institution{Huawei}
  \country{China}}

\renewcommand{\shortauthors}{Wang et al.}
\authorsaddresses{Luming Wang, Xu Zhang, Songyue Wang, Zhuolun Jiang, Tianyue Lu, Mingyu Chen, \{wangluming, zhangxu19s, wangsongyue22s, jiangzhuolun22z, lutianyue, cmy\}@ict.ac.cn, Institute of Computing Technology, Chinese Academy of Sciences, Beijing, China
and School of Computer Science and Technology, University of Chinese Academy of Sciencess, Beijing, China; Siwei Luo, Keji Huang, \{luosiwei, huangkeji\}@huawei.com, Huawei, China}

\begin{abstract}
The growing memory demands of modern applications have driven the adoption of far memory technologies in data centers to provide cost-effective, high-capacity memory solutions. However, far memory presents new performance challenges because its access latencies are significantly longer and more variable than local DRAM. For applications to achieve acceptable performance on far memory, a high degree of memory-level parallelism (MLP) is needed to tolerate the long access latency.

While modern out-of-order processors are capable of exploiting a certain degree of MLP, they are constrained by resource limitations and hardware complexity. The key obstacle is the synchronous memory access semantics of traditional load/store instructions, which occupy critical hardware resources for a long time. The longer far memory latencies exacerbate this limitation.


This paper proposes a set of Asynchronous Memory Access Instructions (AMI) and its supporting function unit, Asynchronous Memory Access Unit (AMU), inside contemporary Out-of-Order Core. 
AMI separates memory request issuing from response handling to reduce resource occupation.
Additionally, AMU architecture supports up to several hundreds of asynchronous memory requests through re-purposing a portion of L2 Cache as scratchpad memory (SPM) to provide sufficient temporal storage. Together with a coroutine-based programming framework, this scheme can achieve significantly higher MLP for hiding far memory latencies.

Evaluation with a cycle-accurate simulation shows AMI achieves 2.42$\times$ speedup on average for memory-bound benchmarks with 1$\mu$s additional far memory latency. Over 130 outstanding requests are supported with 26.86$\times$ speedup for GUPS (random access) with 5 $\mu$s latency. These demonstrate how the techniques tackle far memory performance impacts through explicit MLP expression and latency adaptation.
\end{abstract}




\maketitle

\section{Introduction\label{sec:intro}}
In recent years, the demand for memory has grown rapidly due to the prevalence of big data applications, such as in-memory databases and graph processing \cite{lagar2019software}. However, the slowdown of DRAM device scaling \cite{Lee2016Technology} has constrained the capacity of local memory. As a result, various far memory techniques such as CXL-based memory expansion, disaggregated memory, and Non-Volatile Main Memory (NVMM) have emerged to help address the rising demand for larger and cost-effective memory capacities. In this paper, far memory refers to all alternative memory technologies that provide higher capacity while maintaining standard direct load/store access semantics used by applications. This allows memory-hungry workloads to extend the size of the working set easily.  

\begin{figure}[tb]
  \centering
  \includegraphics[width=0.5\linewidth]{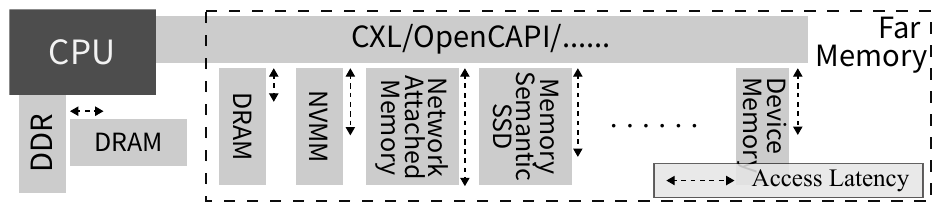}
  \caption{Far Memory in Emerging Data Center Servers\label{fig:future-server}}
\end{figure}

While providing high capacity at low cost, far memory also introduces latencies that are significantly longer and more variable than local DRAM. In a system with heterogeneous far memory devices (Figure \ref{fig:future-server}), the latencies may range from 200ns to over 5$\mu$s. It will bring considerable challenges to performance optimization, as applications and modern processors have been highly optimized based on the assumption of DRAM latency. 

Traditional techniques for tolerating latency include caching, bulk data transfer, and memory access overlapping. Due to the poor temporal and spatial locality of many big data workloads, the first two methods are inadequate normally. On the other hand, big data programs often exhibit a large number of independent memory operations. Therefore, by overlapping more memory requests, achieving a high degree of memory-level parallelism (MLP) is the primary solution to effectively hide the latency of far memory. 

Some domain-specific accelerators or many-core processors are capable of exploiting high MLP by providing a large number of lightweight hardware threads. However, these solutions are typically designed for specific applications and unsuitable for data processing with complex control flow. Meanwhile, general-purpose out-of-order (OoO) processors remain the mainstream in data centers due to their balanced cost and performance. With far memory adoption on the rise, the achievable MLP of OoO processors takes on increasing importance for overall performance.

However, the MLP that modern OoO processors can achieve is limited by their restricted instruction window, which is insufficient to hide the latency of far memory. OoO cores use complex hardware structures and logic like the reorder buffer (ROB), load/store queue (LSQ), and miss status holding registers (MSHRs) to extract potential MLP and track outstanding memory requests. Hence, instruction windows are highly constrained by hardware. Meanwhile, cache-missed memory operations can hold certain hardware resources for a long time, easily leading to resource exhaustion and pipeline stall. 

\begin{figure}[tb]
\centering
\includegraphics[width=0.73\columnwidth]{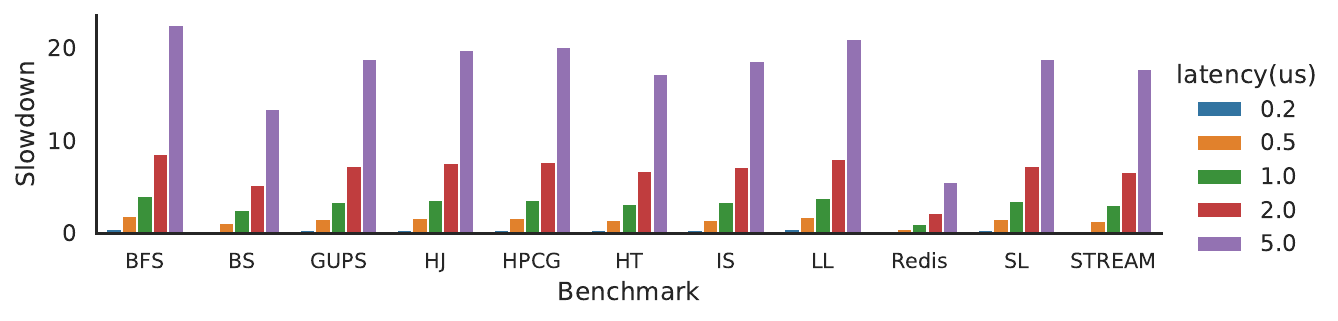}
\caption{Benchmark slowdown under different far memory latencies.
The Y-axis shows the normalized performance slowdown relative to the performance under 100 ns far memory access latency.\label{fig:bench-slowdown}}
\end{figure}

The even longer latencies of far memory exacerbate this issue. Memory-bound workloads operating on far memory experience significant performance degradation compared to utilizing local DRAM. As Figure \ref{fig:bench-slowdown} shows, typical memory-bound workloads experience 3-4x slowdowns when far memory latency increases to 1 $\mu$s, which is state-of-the-art network latency \cite{smartnic}.

While increasing hardware resources such as the ROB and MSHRs can improve MLP, the scalability of these resources is limited by their complex hardware logic \cite{Alipour2019}. Significant prior works have aimed to overcome these scalability issues \cite{srinivasan2004continual,tuck2006scalable,asiatici2019stop,litzCRISPCriticalSlice2022, naithani2021vre} in the past decades. However, the out-of-order windows of current state-of-the-art commercial processors are still limited to the order of several hundred. Achieving sufficient MLP for hiding the far memory latency still requires several times of that scale. Thus, focusing solely on increasing hardware resources is inadequate.


The key factor contributing to this problem is that the \textbf{synchronous semantics} of traditional load/store instructions cause them to occupy critical hardware resources for a longer time when facing long latencies. This prevents other independent memory operations inherent in programs from being issued further, restricting the maximum MLP that can be achieved. 

Addressing this challenge necessitates asynchronous memory access techniques that separate memory request issuing from memory response. 
Instructions that invoke an asynchronous memory request can be retired right after issuing the request, leading to the immediate release of associated hardware resources rather than occupying resources for a full access period, as with traditional synchronous load/store instructions.

Prefetching and external memory access engines are two typical asynchronous mechanisms in modern processors. However, existing asynchronous memory access techniques still have limitations for far memory scenarios. While prefetching enables asynchronous memory request invoking, it lacks support for tracking request completion and managing returned data, limiting its effectiveness for complex scenarios with highly varying far memory latencies. Meanwhile, offloading to external memory engines incurs high startup and notification overhead, hindering its applicability to fine-grained and irregular memory accesses common in big data applications such as graph computing and in-memory databases. 
In a word, there lacks a better asynchronous mechanism inside OoO core to achieve high MLP when facing the issues of long and variable far memory latencies.

This paper proposes a set of novel Asynchronous Memory Access Instructions (AMI) and an Asynchronous Memory Access Unit (AMU) inside a modern OoO CPU core. AMI offers asynchronous initiation and notification mechanisms to reduce critical resource occupation during accesses and enable full software management flexibility. The AMU provides efficient support for the AMI as well as management for massive outstanding requests in a cost-efficient manner. 

The AMU achieves low-cost hardware resources and efficient execution of AMI through its tight integration within the processor core. It dynamically reserves a portion of the L2 cache as Scratchpad Memory (SPM). 
The SPM serves two critical functions. First, it provides program-managed data storage to address the register pressure issues. Second, the SPM offers sufficient state storage to track outstanding requests without costly content-addressable memory (CAM)-based queues. 

AMI introduces \textit{aload} and \textit{astore} instructions for invoking asynchronous memory access requests and \textit{getfin} for polling asynchronous responses. All instructions of AMI do not hold any general registers to avoid pipeline stalls. In fact, AMIs only move data between SPM and far memory, leaving the normal synchronous \textit{load} and \textit{store} to move data between registers and SPM. 

While using L2-SPM instead of various registers enables scaling up easily, three new performance issues must be solved: The first is that metadata access latency changes from register level to L2 level. The second is that when pipeline roll-back occurs due to failed speculation, state consistency in L2-SPM must also be maintained. The third is how to support memory disambiguation for massive outstanding memory operations without CAM-based hardware support. In this paper, we also present our correspondent solutions as metadata caching, AMU speculation, and software  disambiguation. 
 

To address the programming complexity introduced by AMI, we also develop a coroutine-based framework implemented in C++. This framework abstracts away the low-level details of instruction scheduling and SPM management, providing a simplified programming model. 

Overall, we present a hardware-software cooperative solution to achieve high MLP for OoO processor core, allowing for efficient utilization of unlimited far memory resources.

Our contribution can be concluded as follows
\footnote{This paper is an original work based on our previous work \cite{WANG2022100061, wangArchitectureRISCVISA2022} which proposes the basic concept of AMU as well as an implementation on an in-order core. This paper provides a novel design and implementation of the AMU on an OoO processor core, an in-depth discussion and software solution for addressing data conflicts introduced by the AMU, details of the programming framework and compiler support for leveraging AMU, and a more thorough evaluation characterizing performance, power consumption, software overhead, and hardware resource utilization.}:
\begin{itemize}
  \item A set of novel Asynchronous Memory Access Instructions (AMI) that model memory requests and responses as separate instructions. This reduces critical resource occupation and exposes more opportunities for  software to exploit MLP.
  \item A novel Asynchronous Memory Access Unit (AMU) architecture that efficiently executes AMI. AMU repurposes cache as metadata storage and program-managed data storage to support massive outstanding AMI requests.
  \item An optimized micro-architecture design for AMU, including metadata caching and AMU speculation mechanism, etc.  
  \item An efficient and software-only memory disambiguation mechanism to substitute the non-scalable CAM hardware. 
  \item A coroutine-based C++ framework to address programming complexity from AMI. The framework encapsulates low-level scheduling and SPM management.
  \item A cycle-accurate model of AMU is built based on the Gem5 simulator. The evaluation results show that for memory-bound benchmarks, AMU achieves an average speedup of 2.42$\times$ when the additional latency introduced by far memory is 1 $\mu$s. In the case of the random access benchmark from HPCC, our technique can offer a speedup of 26.86 $\times$ when the far memory latency is 5 $\mu$s, with the average number of in-flight memory requests of single-core exceeding 130.
\end{itemize}

\section{Background and Motivation\label{sec:motivation}}
\subsection{Far Memory and MLP}
Far memory refers to a variety of memory technologies that offer higher capacity compared to DRAM, including device memory/remote memory based on Non-Volatile Main Memory (NVMM) \cite{liu_survey_2021}, cache-coherent interconnect-based memory/devices (e.g. CXL \cite{sharmaComputeExpressLink2022}), and Disaggregated Memory. While offering much larger memory capacities, far memory presents challenges because of its long and highly variable latency. 


\textbf{Non-volatile Main Memory (NVMM)} provides higher storage density and persistence by using new materials \cite{inteloptane, sttram2013, Qureshi2009ScalableHP}. Using as main memory, NVMM shows a higher access latency than conventional DRAM. One example is the Intel Optane DC Persistent Memory Module, which exhibits an access latency that ranges from 200 to 300 nanoseconds \cite{liu_survey_2021}. Another example is ultra-low latency flash (ULL-Flash) based devices \cite{2018Exploring}, which achieve a typical read latency of 3$\mu$s and write latency of 100$\mu$s. To date, NVMM has seen limited commercial adoption while various researches are still underway.

\textbf{Cache-Coherent Interconnect Based Memory/Devices} enables CPUs to directly access memory and storage devices without the overhead of extra software mechanisms like page swapping. This is achieved through emerging coherent interconnects such as Compute Express Link (CXL) \cite{sharmaComputeExpressLink2022}, Open Coherent Accelerator Processor Interface (OpenCAPI) \cite{openCAPI} and Gen-Z \cite{GenZ}. Direct attached CXL-enabled memory has a typical latency of approximately 200ns \cite{sharmaComputeExpressLink2022}, which is still 3-fold larger than conventional local DRAM. For memory-semantic SSD, the access latency can be 35\%-738\% slower than local DRAM, depending on the contention of SSD \cite{kwonCacheHandExpanderDriven2023}. CXL-enabled memory extensions have gotten more attention in recent years. Anyway, local extension within a node still has its capacity limitations.



\textbf{Disaggregated Memory} allows programs to directly leverage memory resources on remote nodes based on high speed interconnects. Both academia \cite{liPondCXLBasedMemory2022a} and industry \cite{IBMPower10} have developed remote memory prototypes and products based on emerging interconnects. For coming CXL-enabled disaggregated memory systems, the latency depends on propagation delays, including the latency of switches and ports, which will reach to more than 300ns \cite{liPondCXLBasedMemory2022a}. For network-attached remote memory across multiple switches, the latency can reach 1$\mu$s-2$\mu$s, which is the state-of-the-art record \cite{smartnic}. Furthermore, the newly released CXL 3.0 specification introduces support for Global Fabric Attached Memory (GFAM), enabling the construction of memory pools that across multiple nodes. 

Besides different types of far memory technologies, highly variable latencies also come from other factors. 
Due to queueing latency and contention, accessing the same level of the memory hierarchy can exhibit noticeable latency variations \cite{diavastosEfficientInstructionScheduling2022}. The potentially complex hierarchical structures inside far memory devices can also cause significant differences in memory access latencies.

Long and highly variable latencies further widen the gap between processor and memory. This exacerbates the traditional memory wall problem. Therefore, applications that want to get the benefit from the large capacity of far memory have to seek ways for latency tolerance first.

Traditional techniques for tolerating latency include caching, which reduces average access latency by keeping hot data locally, and bulk data transfers, which move large contiguous blocks to minimize transaction count. However, both methods require memory locality. Big data applications often have extremely large working sets that are not cache-friendly. Some big data applications, such as graph computing and in-memory databases,
exhibit fine-grained random patterns with poor temporal and spatial locality. Consequently, adjacent memory locations may not be related, making direct large block data transfers inefficient \cite{10188866}.

Fortunately, these workloads often comprise numerous  memory operations with weak dependencies, allowing potential opportunities for MLP extraction. Yet mechanisms are needed to fully exploit the potential MLP.




\subsection{The limitation of Out-of-order CPU for MLP \label{subsec:hardware-latency-tolerant}}
The traditional approach to extracting MLP from workloads relies on aggressive OoO execution of CPU cores. By providing a wider instruction window and leveraging aggressive OoO execution, cores can issue more in-flight load/store instructions for higher MLP.

The limitation of OoO execution is that it relies on complex hardware logic to track in-flight instructions and memory requests, resulting in poor scalability. For instance, structures like the LSQ and MSHRs are typically built using CAM, and increasing CAM capacity raises power consumption and latency rapidly. While significant past works \cite{srinivasan2004continual,tuck2006scalable,asiatici2019stop,litzCRISPCriticalSlice2022, naithani2021vre} targeted these scalability issues, merely expanding hardware are cost-ineffective. The MLP achievable with OoO execution is still constrained by certain critical hardware resources.

To overcome these constraints, previous research explores some approaches to utilize existing resources better to improve OoO aggressiveness, such as Runahead Execution \cite{mutlu2003runahead, hashemi2015, naithani2020pre, naithani2021vre, naithani2023dvre} and CLEAR\cite{kirmanCheckpointedEarlyLoad2005a}. Runahead Execution allows processors to enter a "Runahead" mode when the ROB is exhausted, speculatively executing subsequent instructions to prefetch future memory accesses. CLEAR speculatively retires long-latency loads early to free up resources for other instructions. However, these kinds of techniques only address the issue of individual resource shortages like the ROB, but cannot handle the exhaustion of other resources such as MSHRs. As a result, they introduce more complex hardware logic without fundamentally solving the scalability issue for far memory.



The MLP demand for far memory has significantly exceeded the potential of contemporary OoO CPU. Assuming a far memory access latency of 1$\mu$s and a processor main frequency of 2GHz, the access latency of a single far memory equates to 2000 cycles. 
As memory operations comprise ~30\% of instructions \cite{limayeWorkloadCharacterizationSPEC2018}, avoiding all stalls would require thousands of ROB entries and hundreds of MSHRs. This is impractical for current technology. State-of-the-art processor cores only have several hundreds of ROB entries and several tens of MSHRs.

The key reason here is, in fact, the synchronous semantics of traditional load/store instructions. A synchronous operation occupies critical pipeline resources until it finishes. Due to the significantly longer latencies of far memory, processor resources are exhausted more rapidly, leading to more frequent pipeline stalls and preventing further memory operations from being dispatched.


\subsection{Asynchronous Memory Access Techniques\label{subsec:async_memacc_techs}}
To address the resource occupation issue, asynchronous memory accesses that separate memory request issuing from memory response processing are needed. 
Asynchronous memory accesses allow corresponding instructions to be committed once the memory request is issued, instead of blocking for the response. Thus, resources like ROB entries and physical registers can be freed earlier rather than occupying them for the full latency duration. Additionally, asynchronous memory access also enables software to be further involved in exposing more opportunities for MLP extraction. 


\textbf{Prefetching} is a typical and widely used asynchronous memory access technique\cite{falsafiPrimerHardwarePrefetching2014,yuIMPIndirectMemory2015,ishiiAccessMapPattern2009,bakhshalipourDominoTemporalData2018,bakhshalipourBingoSpatialData2019,chen2007improving,kocberber2015asynchronous, tranSWOOPSoftwarehardwareCodesign2018}, which can be considered as a form of asynchronous load. Prefetch instructions do not hold ROB resources after the load request is issued. But prefetch loads still occupy MSHRs until data are returned later and put in cache for best effort. 
As an asynchronous mechanism, prefetching commonly faces timeliness issues \cite{leeWhenPrefetchingWorks2012}. The prefetched data can either be evicted from caches before use (i.e. early prefetches) or arrive after the demand loads have occurred (i.e. late prefetches). Thus, the benefits of prefetching are limited without sophisticated optimization.

\begin{figure}[tb]
  \setlength{\abovecaptionskip}{0pt} 
  \setlength{\belowcaptionskip}{-1pt}
\includegraphics[width=0.78\columnwidth]{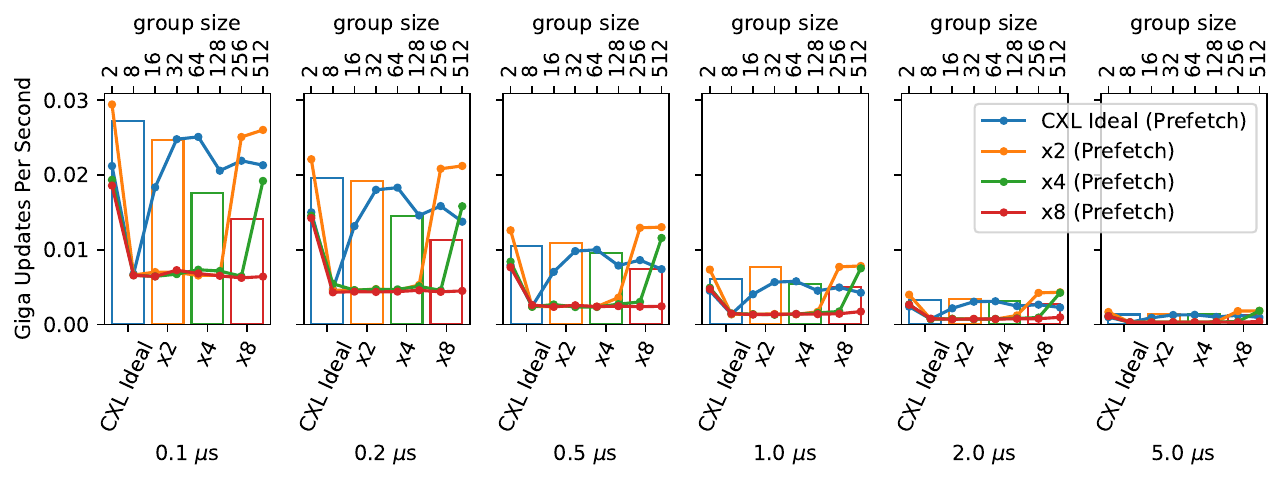}
\centering
\caption{The performance of GUPS as hardware resources are increased\label{fig:gups_prefetch_motivation}. The \textbf{CXL Ideal} configuration is described in Section \ref{sec:evaluation}. The hardware resources (i.e. IQ, LSQ, ROB, MSHRs, and physical registers) of \textbf{x2} and \textbf{x4} are two times and four times those of \textbf{CXL Ideal} configuration, respectively. The lines show the performance of Group Prefetching \cite{chen2007improving} based GUPS under different group (or batch) sizes. The bar plot represents the performance of the baseline GUPS benchmark.}
\Description{}
\end{figure}

Figure \ref{fig:gups_prefetch_motivation} demonstrates this weakness, comparing the performance of a GUPS benchmark using GP\cite{chen2007improving}-based prefetching to a baseline. GUPS involves random memory updates. The GP-based GUPS variant prefetches all addresses to be updated in a group before executing each update. As shown in Figure \ref{fig:gups_prefetch_motivation}, the effectiveness of GP heavily depends on the group size. different group sizes can cause the GP-based GUPS to outperform or underperform the baseline on the same hardware configuration and latency conditions. Moreover, the group size yielding the best performance varies greatly with hardware configurations. This example shows that prefetching would struggle to adapt to unpredictable latencies.

The limitation of prefetching is caused mainly by a lack of support for tracking responses and managing returned data. Software cannot obtain information about request completion, making it difficult to ensure that the prefetched data is available when needed. As a result, prefetching is not well-suited for scenarios with far memories that have highly variable latencies. 



\textbf{External memory engines} also provide asynchronous memory access via offloading expensive memory operations to external engines for asynchronous execution. With an I/O-like interface, it provides response notification and explicit management of data storage. Typical techniques include Intel I/O Acceleration Technology (I/OAT), and Intel Data Streaming Accelerator. 

However, these engines face high startup overhead. They generally require setting up and en-queuing descriptors for the engine to initiate memory requests. Although advanced engines allow directly converting writes into remote memory requests (e.g. Cray FMA), this still requires several-tens cycles as the engines are attached to NoC or I/O bus. Therefore, this approach is typically only suitable for applications that involve transferring large, contiguous blocks of data, but not applications with fine-grained and random accesses.

To reduce overhead, some works \cite{morariScalingIrregularApplications2014,nelsonLatencyTolerantSoftwareDistributed2015} have proposed aggregating many small messages issued by user-level threads into a larger packet to lower the average overhead. However, the software-based aggregation increases the latency of individual memory operations, necessitating applications to have inherently high MLP to fully hide the increased latencies.

Additionally, Some works \cite{corbalCommandVectorMemory1998, orenes-veraTinyMightyDesigning2022} have proposed programmable memory engines. These engines are also located on the NoC, allowing complex memory access patterns through compiler or manual programming. However, these engines typically adopt a produce/consume model for interaction with the host core. The host core will send a non-speculative notification signal to the memory engine after consuming the data. These synchronous operations eliminate the advantages of out-of-order cores with aggressive speculative execution. Consequently, such works are more suitable for processors with simple cores such as in-order cores, VLIW, or many-core processors with many small cores rather than high-performance out-of-order designs.

Integrating external memory access engines inside the core is not straightforward. The engines would need to be redesigned to support the cancellation of requests for handling mis-speculation. 
Relying on the existing accelerator interfaces like Rocket Chip Coprocessor (RoCC) \cite{Asanović:EECS-2016-17} is also not possible. Because they lack speculative execution support, which makes them unsuitable for implementing tightly coupled coprocessors \cite{masterthesis:2016-2017}.
Therefore, directly integrating external memory engines into an OoO pipeline presents difficulties.


In summary, existing asynchronous techniques still have limitations in far memory scenarios. Prefetching lacks a notification mechanism, making it difficult to apply to far memories with highly variable latencies. Offloading to external memory access engines incurs high startup overhead, making it challenging to apply to big data applications involving word-sized random memory access patterns, such as graph computing workloads.

\subsection{Motivation \label{subsec:motivation}}



In this paper, we want to design a built-in asynchronous memory access mechanism for an OoO core to meet the MLP demand of far memory without sacrificing the performance of local operations. Based on the previous discussions, an asynchronous memory access technique targeted far memory scenarios should support asynchronous request issuing, response notifications, and explicit storage management. 

Our first design choice is where to put the storage. To meet the typical demand of far memory, up to several hundreds of simultaneous memory operations should be supported for a single OoO core. The data and metadata needed for it require tens of KBs of space. It exceeds the existing scope of register files even L1 cache. So, we choose to dynamically partition a portion of the L2 cache as SPM for massive far memory operations. While L2 partitioning incurs some impact, this is acceptable versus the performance gains of AMU. The dynamically software-adjustable SPM size avoids affecting non-AMU applications. Additionally, big data applications typically underutilize L2 caches \cite{zhangMakingCachesWork2017, leeMERCIEfficientEmbedding2021}. Occupying part of the L2 cache as SPM thus has limited negative performance impact. 

Secondly, we propose a set of new asynchronous instructions for accessing far memory, both for request initiation and response notification. The key design rule is that an asynchronous instruction should not hold any registers for long. Once a memory request is sent out, the initiation instruction should be retired at once. The later response check instruction should not be in a block-and-wait mode either. Between the two instructions, data and metadata for a pending memory operation are stored in the SPM rather than the hardware registers. As register allocation is done by modern compilers, we do not use hardware instructions for SPM data allocation and leave it for software. The new asynchronous instructions are only in charge of moving data between SPM and far memory. Moving data between GPR and SPM is done by normal synchronous \textit{load/store} instruction. These operations then will have short and fixed latency since no cache-miss will occur anymore. Asynchronous and synchronous instructions work cooperatively for the high MLP. 

Thirdly, we propose adding a special function unit for the new asynchronous instructions. The new asynchronous instructions can share the same fetching, decoding, and dispatching stage as normal instructions in an OoO pipeline. However, since asynchronous instructions have rather different requirements for resource scheduling and interact heavily with SPM, we propose to add a new function unit for their execution phase. To some extent, it is like a vector unit for vector instructions and vector registers.

Fourth, we choose to handle in-thread data consistency through software instead of non-scalable hardware. There are two possible sources of data conflicts and inconsistencies for instruction sequences. One is between asynchronous memory accesses and traditional load/store instructions. The other is conflicts among asynchronous instructions themselves. For consistency between load/store and asynchronous instructions, we choose to use dynamic partition to avoid hardware complexity. By partitioning, we could assume that asynchronous instructions and load/store would not access the same memory region simultaneously. 
Necessary cache flush operations are needed for region transition.

For consistency among asynchronous instructions themselves, it means disambiguation for these new memory access operations. Hardware handling disambiguation would have too high costs due to the large number of outstanding asynchronous memory requests. On the other side, big data programs often employ data parallelism or have large datasets, so the probability of data conflicts is low. Therefore, we choose to do conflict detection through software when necessary. For efficiency, data structures for disambiguation can be held either in SPM or local cache with flexibility.

We name the new asynchronous instructions and the supporting functional unit as AMI and AMU. Together with AMI and AMU, software can get full explicit control over massive memory requests to exploit enough MLP for far memory accesses. 

\section{Overview of AMI and AMU\label{subsec:key_idea}}
\begin{figure}[tb]
  \setlength{\abovecaptionskip}{0pt} 
  \setlength{\belowcaptionskip}{-1pt}
\includegraphics[width=0.88\columnwidth]{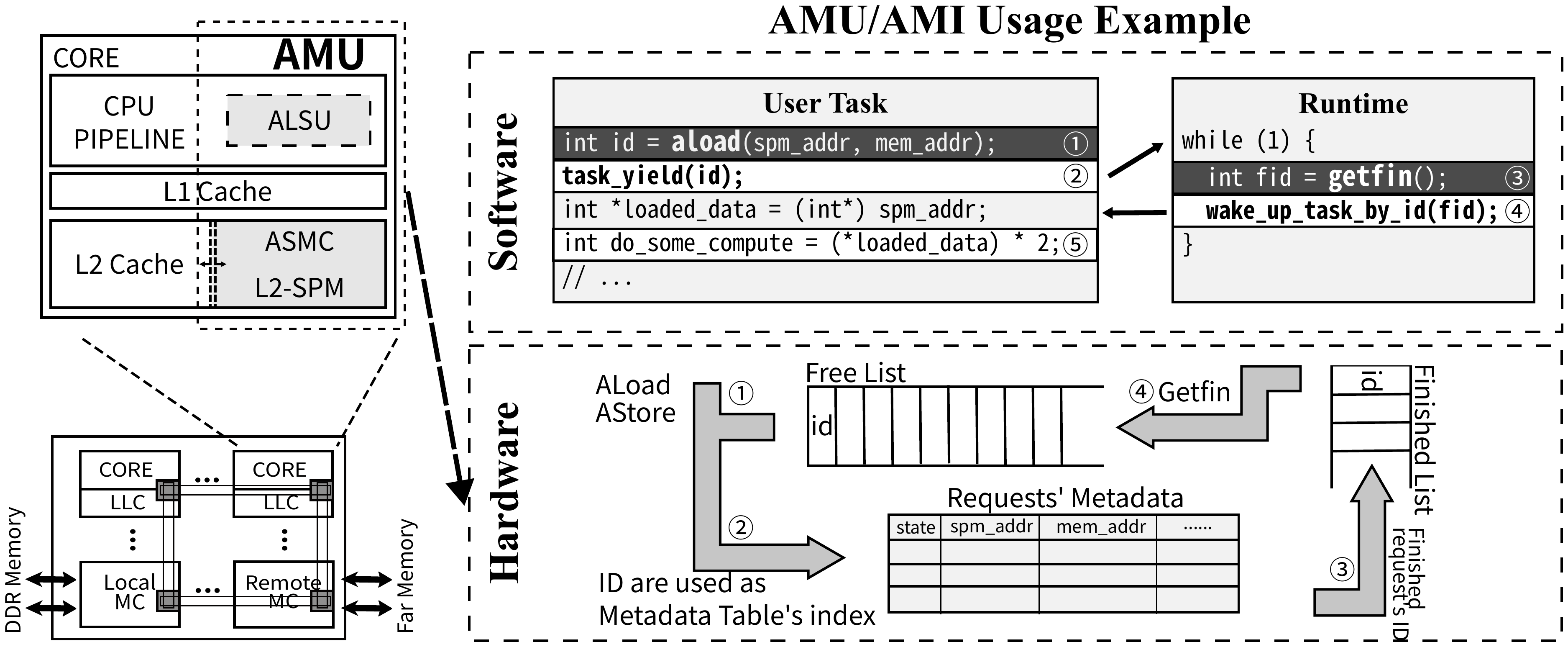}
\centering
\caption{AMU Architecture \label{fig:arch-sim-overview}}
\end{figure}

Figure \ref{fig:arch-sim-overview} shows the overall architecture of AMU as well as an example to illustrate the programming model of AMI. The design is based on our previous studies \cite{WANG2022100061, wangArchitectureRISCVISA2022}, which only provided a preliminary discussion of AMI and an in-order core AMU prototype without investigation into the challenges of OoO-based design. 

\subsection{Asynchronous Memory Access Instruction and The Programming Model\label{subsec:programming_model}}
AMI contains three basic instructions (listed in Table \ref{table:programming_interface}). \textit{aload}/\textit{astore} instructions are designed to invoke asynchronous data movement requests between SPM and memory. The \textit{getfin} instruction, inspired by I/O multiplexing, returns the ID of a completed request to serve as a notification mechanism. Conventional hardware mechanisms for notification are interrupt-based or polling-based, but these are unsuitable in this case. Interrupt-driven notifications incur too much overhead for memory requests, while polling-based schemes waste cycles repeatedly checking the status of each request. In contrast, the \textit{getfin} enables software-based out-of-order scheduling of instruction streams or tasks. The scheduling is dependent on memory access completions.

From the software perspective, the SPM acts as a special type of register. The AMI provides the ability to load/store contents of far memory to/from the SPM. After software initiates an asynchronous memory access, the AMU copies data between far memory and the SPM. Subsequently, software operates the data in SPM through standard load/store instructions.

The configuration registers supported by AMI are listed in Table \ref{table:programming_interface}. These registers are configured by executing \textit{cfgrr}/\textit{cfgrw} instructions. Applications can configure the start location (i.e. \textit{queue\_base}) and the range of metadata area in SPM. The maximum number of outstanding memory requests it requires is controlled by \textit{queue\_length}. AMU will initialize the specified number of metadata table entries to be used as metadata area. Then, the SPM metadata area is managed by AMU only. Other areas of SPM are used as data storage visible to instructions. 

A brief usage example of AMI is shown in Figure \ref{fig:arch-sim-overview}. The example assumes the software employs lightweight threads for dynamic instruction stream scheduling. \ding{172} A user task initiates an asynchronous memory request using the \textit{aload} instruction, which returns an identifier representing the request. \ding{173} The user task suspends and notifies the runtime system of the request ID it is waiting on. \ding{174} The runtime system executes a \textit{getfin} instruction in an event loop to obtain the ID of a completed request. \ding{175} The runtime system then awakens the user task waiting on that ID. \ding{176} The user task accesses the retrieved data in the SPM using a standard \textit{load} instruction.

Additionally, the proposed AMI supports variable granularity access (through configuring the granularity of memory access requests), which naturally reduces the algorithm design overhead in several domains (e.g. graph processing \cite{zhangMakingCachesWork2017}). Software does not need to use multiple instructions to combine or split data to match register or cache line size. A single \textit{aload} can access large data blocks up to KBs.

For address translation, the virtual addresses in \textit{aload}/\textit{astore} are translated to physical addresses via the traditional TLB. Address translation in distributed memory systems is an active research area \cite{hajinazarVirtualBlockInterface2020} but orthogonal to this work. Therefore, we do not consider address translation impacts and assume a conventional address translation. Also, the SPM uses a fixed mapping and does not require address translation.

However, the asynchronous programming model of AMI increases program complexity versus conventional synchronous models. Approaches are required to reduce this complexity and lower the programming effort for developers. Section \ref{sec:programming_paradigm} discusses programming frameworks as an approach to tackle programming complexity.

\begin{table}[tb]
  \setlength{\abovecaptionskip}{0pt} 
  \setlength{\belowcaptionskip}{-1pt}
  \centering
    \footnotesize
  \begin{tabular}{lp{13cm}}
    \toprule
    \textbf{Instruction} & \textbf{description}\\
    \midrule
    \begin{tabular}[t]{@{}l@{}}\textit{aload Rd, Rs1, Rs2}\\ \textit{astore Rd, Rs1, Rs2} \end{tabular} & Initiate an asynchronous memory access request to read the data from \textit{Rs2 (Memory Address)} to \textit{Rs1 (SPM Address)} or write the data from \textit{Rs1 (SPM Address)} to \textit{Rs2 (Memory Address)}. The returned request ID is placed in \textit{Rd}. If the ID allocation fails, \textit{Rd} is set to 0. \\ 
    \addlinespace[0.1cm]
    \textit{getfin Rd} & Get a completed ID of request. If there is no finished request, the instruction returns a failure code. \\
    \midrule
    \textit{cfgrr rd, CFGREG} & Read the specified configuration register \textit{CFGREG} to the general-purpose register \textit{Rd}. \\
    \textit{cfgrw rs1, CFGREG} & Writes the value of the general-purpose register \textit{Rs1} to the configuration register \textit{CFGREG}. \\
    \toprule
    \textbf{Registers} & \textbf{description} \\
    \midrule
    \textit{granularity} & Configuring the granularity of memory access requests \\
    \textit{queue\_base} & Configuration of the SPM metadata area start address \\
    \textit{queue\_length} & Configuration of the length of the SPM metadata area \\
    \bottomrule
  \end{tabular}
  \caption{Basic Asynchronous Instructions And Configuration Registors}
  \label{table:programming_interface}
\end{table}

\subsection{Asynchronous Memory Access Unit}
AMU is a memory access accelerator integrated within processor cores. When software issues requests, the AMU is responsible for converting them to native memory requests and tracking their status. These requests are sent to the remote memory controller (MC) through the local bus. The remote MC communicates with the far MC through a high-speed network or I/O bus. This paper will not discuss the details of remote MC and far MC. 

The AMU (Figure \ref{fig:arch-sim-overview}) consists of two main components: the Asynchronous Load/Store Unit (ALSU) integrated into the pipeline and the Asynchronous Scratchpad Memory Controller (ASMC) integrated with the L2 cache controller. The ALSU is specifically responsible for executing AMI instructions to generate the corresponding asynchronous memory requests. These requests are sent to the ASMC via the standard ports of the cache. The ASMC is responsible for managing the SPM and converting ALSU requests into far memory requests. It is integrated with the L2 cache controller.

The procedure of managing asynchronous requests is shown in Figure \ref{fig:arch-sim-overview}. \ding{172} When a user executes an \textit{aload}/\textit{astore} instruction, the AMU allocates a free ID from the free list for the current asynchronous memory request. The ID is returned to the user program. \ding{173} Then, AMU maintains the corresponding list entry and metadata in the SPM, using the ID as the index for the metadata table. After that, AMU schedules an un-core memory access to remote MC. The \textit{aload}/\textit{astore} instructions can be retired at this point. Meanwhile, the corresponding LSQ and ROB resources are released. \ding{174} When the request is completed, AMU places the ID into the finished list. \ding{175} Applications retrieve completed IDs using the \textit{getfin} instruction. \textit{getfin} returns the ID and puts it back into the free list for subsequent requests. The maximum number of in-flight memory accesses depends on the size of the SPM meta-data area, making even hundreds-level MLPs supported easily.


The AMU repurposes on-core L2 cache resources as SPM to provide both data storage and a request metadata store. This avoids the register pressure \cite{tranSWOOPSoftwarehardwareCodesign2018} and MSHRs constraints faced by memory-intensive applications \cite{zhangMakingCachesWork2017} when following asynchronous memory access paradigms. This enables efficient tracking of several hundreds of outstanding requests and helps scale the AMU design with modest hardware overhead.




Additionally, AMU takes advantage of SPM to enable variable granularity memory access. AMU can support large granularity memory access through instructions rather than traditional I/O interfaces. In contrast, designs relying solely on registers are limited by register size and can only load data at the granularity of individual registers. This makes it difficult to support asynchronous access to a large block of data in a single instruction. SPM allows data to be loaded in chunks of varying sizes, from the smallest units up to very large blocks without restriction by register size. This approach provides opportunities to optimize performance, as programs can efficiently load and process data with granularities best suited to the computational problems. Furthermore, when programs require loading a large block of data, this avoids the latency and overhead of loading smaller chunks individually through separate memory accesses.

However, the AMU presents three challenges that must be addressed.

First, while using an SPM to replace physical register files and MSHRs provide greater capacity, SPM accesses are significantly slower. Mechanisms are needed to mitigate this performance gap. Section \ref{subsec:batching_id} investigates micro-architectural optimizations to overcome the speed disadvantage of accessing data stored in the SPM.

Second, supporting speculative OoO execution requires designing rollback mechanisms for AMU. Although speculatively execution of traditional load instructions affects cache state, there are no program-visible side effects. In contrast, AMI instructions have program-visible side effects as they modify the state and data stored in the SPM. The SPM-based design of AMU therefore complicates the rollback process since it involves restoring SPM contents when misspeculation occurs. This poses additional challenges compared to conventional cache-based designs in undoing misspeculation effects. Section \ref{subsec:amu_ooo} details the speculative execution support of AMU.


Third, SPM lacks associative lookup abilities needed to perform memory disambiguation transparently in hardware, as typically done using CAM-based structures. For instance, OoO cores rely on CAM-based LSQ to implement memory disambiguation. However, CAM structures are expensive to scale. To avoid these limitations, AMU requires a CAM-free approach for memory disambiguation. Section \ref{sec:consistency_checking} examines techniques to address consistency issues through software-based memory disambiguation.


\section{Micro-architecture design\label{sec:architecture_design}}
\subsection{Implementation of the Basic Functionalities}

\begin{figure}[tb]
\includegraphics[width=0.98\columnwidth]{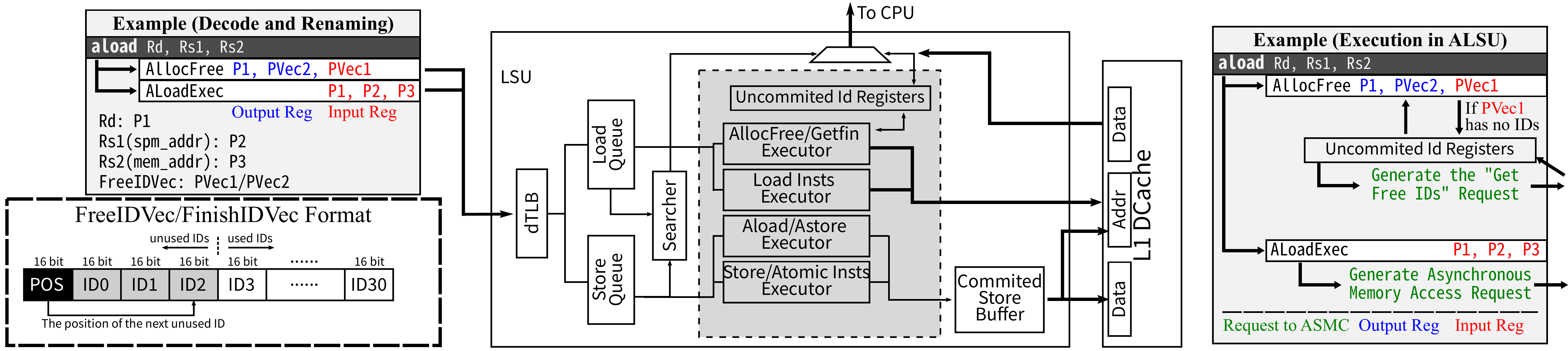}
\centering
\caption{Architecture of ALSU\label{fig:alsu}}
\end{figure}

\begin{figure}[tb]
  \setlength{\abovecaptionskip}{0pt} 
  \setlength{\belowcaptionskip}{-1pt}
\includegraphics[width=0.65\columnwidth]{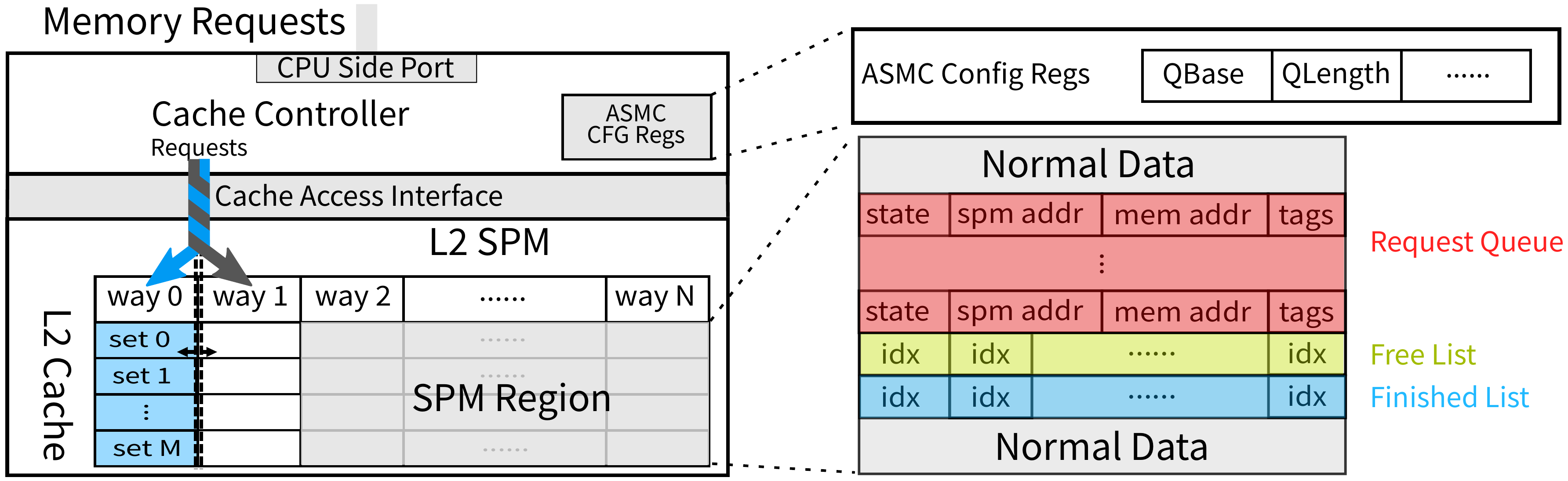}
\centering
\caption{Overview of the ASMC\label{fig:l2-cache-arch}}
\end{figure}


Figure \ref{fig:alsu} presents the structure of ALSU as well as an execution example. The \textit{aload}/\textit{astore} instructions are first decoded into two micro-ops (details in Section \ref{subsec:batching_id}). The first micro-op is used for ID allocation, while the second generates the actual asynchronous memory request. The micro-ops are executed in the ALSU. For ID allocation, the execution unit first checks for available IDs. If a free ID exists, it performs allocation locally. Otherwise, it requests free IDs from the ASMC. After ID allocation, the micro-op for issuing the asynchronous request is executed. The ALSU constructs the asynchronous memory request and passes it to the ASMC. This request is handled similarly to a store request to the cache. The request is buffered in the store buffer before the instruction commit. At this point, the instruction has been completed and can wait in the ROB for retirement.

The ALSU concentrates on the instruction execution and the mis-speculation handling (details in Section \ref{subsec:amu_ooo}). Therefore, the ASMC does not need to consider the intricacies introduced by speculative execution. Communication overhead between the ALSU and ASMC is also reduced through batching of ID transfers (details in Section \ref{subsec:batching_id}).
This lowers the overall design complexity by dividing different concerns between the two units.

Figure \ref{fig:l2-cache-arch} illustrates the design of ASMC. Several modifications are made to the cache controller. First, the control logic is added to repurpose a portion of the cache area as SPM. This implementation is straightforward and is supported by several commercial processors. Second, several new memory commands are defined to support ID-related requests and asynchronous memory access requests. These extensions to the cache controller allow it to serve its original caching functionality while also managing metadata and coordinating memory operations for asynchronous memory access requests. Third, the protocol between L1 cache and L2 cache is expanded to support the newly defined commands. 
Supporting these new commands is straightforward as they do not interfere with maintaining cache consistency.  

The ASMC utilizes three key data structures stored in the SPM metadata area to support the new memory commands: a finished list, a free list, and an Asynchronous Memory Access Request Table (AMART).
For each asynchronous memory request, the ASMC indexes the AMART using a request ID to access the corresponding entry. Each entry contains metadata like the SPM address, memory address, request status, and other implementation-specific information.
The request is then converted to a standard memory request sent to the memory controller. Upon response from memory, the ASMC re-indexes the AMART using the request ID to update the status.
Once an asynchronous request is completed, its ID is written to the finished list by the ASMC. To reduce SPM accesses and improve performance, the ASMC caches a subset of IDs from the finished and free lists in on-chip registers. 

The ASMC enables the transfer of large contiguous data blocks by dividing large requests into cache-line-sized sub-requests. 
The ASMC relies on a dedicated state machine to split large granularity memory requests. The states of memory requests are also tracked in the metadata region of the SPM.



\subsection{Metadata Batching\label{subsec:batching_id}}
\vspace{-5pt}
To reduce overhead from frequent ALSU-ASMC communication, we develop a metadata batching approach.

The key idea involves using vector registers as buffers to aggregate metadata accesses. The major interaction between ALSU and ASMC involves pulling and pushing request IDs from/to the finished and free lists managed by the ASMC. "List vector registers" are thus introduced, holding portions of the IDs in these lists. As shown in Figure \ref{fig:alsu}, each register contains a pointer to the next unused ID entry as well as multiple stored 16-bit IDs. The width of the List Vector Register matches the width of the physical vector registers (512 bits in this paper). 
Thus, the ALSU can request a batch of IDs from the ASMC to store in the list vector register each time, or write back all the IDs in the register with a single request to the ASMC.


 
To minimize software complexity, the list vector registers are not directly accessible to software. Instead, they are accessed through internally-generated micro-ops. Each asynchronous memory access instruction is decoded into two micro-ops: one for performing conditional fetching/writing of IDs between vector registers and ASMC; the other handles the actual functionality of the instruction. Figure \ref{fig:alsu} shows a detailed example. The ID allocation micro-op first checks the list vector register for an available ID. If there are no IDs, it issues a request to the ASMC to fetch a batch of free IDs. Upon receiving the IDs from the ASMC, the allocation continues. 

Figure \ref{fig:alsu} shows the detailed architecture for executing the micro-ops within the ALSU. The process contains two stages: execution and commit.
    
    \textbf{Execution} The micro-ops are sent to the LSQ and dispatched to the corresponding execution unit. Two additional execution units are added to support asynchronous memory access. One unit handles \textit{aload/astore} requests sent to the ASMC, while the other handles ID management micro-operations. The \textit{aload/astore} request is similar to a conventional memory write request with command, data, and address, but with different semantics. The command field specifies \textit{aload/astore}, the data field carries the request identifier and SPM data address encoded in the original instruction, and the address field contains the memory address also from the instruction.
    
\textbf{Commit} The asynchronous memory request is sent to the ASMC when the instruction reaches the head of the ROB, resembling atomic instructions or non-speculative instructions. However, ID management micro-ops can be speculatively executed for higher performance. Their requests may be forwarded to the ASMC prior to the older asynchronous memory access micro-ops.

\subsection{Speculative Execution Support\label{subsec:amu_ooo}}
The primary challenge of supporting speculative execution for AMU is managing the effects of micro-operations that modify metadata, such as finished/free lists in the SPM. For example, if a micro-op is speculatively executed but is later squashed, the speculatively fetched IDs could be lost prematurely. There are three cases of how AMI-decoded micro-ops can support out-of-order execution:

The first case is micro-ops responsible for initiating \textit{aload/astore} requests (e.g. the \textit{ALoadExec} micro-op shown in Figure \ref{fig:alsu}). These micro-ops read register data to construct asynchronous memory requests passed to the ASMC. Because asynchronous memory requests can be regarded as a special type of store request, they are handled similarly by buffering in the store buffer before the instruction commit, just like the \textit{store} instruction.

The second case involves ID management micro-ops like retrieving completed IDs or allocating free IDs. When the list vector register still contains IDs, these micro-ops simply move an ID into a general-purpose register without issuing a batch ID fetch request to the ASMC. This allows it to behave like a regular register-to-register instruction, which can thus be handled via the traditional register renaming mechanism.

The third case involves the ID management micro-ops when the list vector register is empty. In this case, these micro-ops will issue a batch ID fetch request to the ASMC, changing the state of the queues in SPM. To address this problem, ALSU uses an "uncommitted ID register" to isolate ID updates from the ASMC. From the perspective of ASMC, the IDs taken by ALSU can be safely removed from the queue in SPM without considering rollback. As shown in Figure \ref{fig:alsu}, the uncommitted ID register keeps the value of the list vector register corresponding to the micro-op that issued the batch ID fetch request. The value in the uncommitted ID register can be regarded as a checkpoint for this request. When a mis-prediction occurs, the register continues holding the IDs retrieved by the canceled instruction. Subsequent micro-ops that issue batch ID fetch requests will obtain IDs from the uncommitted ID register rather than the ASMC. 
This effectively restores the previous obtained IDs to the list vector register. Therefore, IDs fetched from the ASMC are not lost on mispredictions. The uncommitted ID register only being cleared on instruction commit. So, it can only store the result of one batch ID fetch request, requiring a second ID fetch to stall until the previous one is committed. However, this stall is infrequent, as the list vector register can hold 31 IDs before needing refills. 

\section{Software Support}
\subsection{Software-Based Memory Disambiguation\label{sec:consistency_checking}}
For OoO pipeline, solving memory ambiguity is crucial to ensure program correctness. Specifically, consistency must be maintained when two outstanding instructions access the same memory address. Traditionally, memory ambiguities are detected and handled by CAM-based hardware structures. However, AMU relies on large SPM instead of CAM-based hardware structures to enable high MLP. As a result, memory disambiguation has to be done through software. This section details our software solutions for handling data consistency among AMIs, while handling potential data conflicts between AMIs and \textit{load}/\textit{store} instructions is discussed in Section \ref{subsec:work_with_legacy_code}.


We found that software-based memory disambiguation can be sufficient for two reasons. First, a study has shown that on average 43\% of memory operations do not cause any violations \cite{huangSoftwarehardwareCooperativeMemory2006}. For big data workloads that enjoy far memory, this percentage can be higher as they typically employ data parallelism. Therefore, software memory disambiguation only needs to be applied for a small portion of memory operations. 
Second, only a small fraction of memory locations are active at any given time \cite{zhuSynchronizationStateBuffer2007}. As a result, there is no need for the software to track the status of all memory locations that could cause violations. The software simply needs to maintain a record of the active locations at present. 

Based on these observations, we contend that a small cacheable hash set in local DRAM, capable of tracking the addresses of in-flight asynchronous memory requests, can tackle memory aliasing problems. The hash table design used in this work is based on cuckoo hashing but with a slight variation from typical implementations. Specifically, each hash function maps to its separate table rather than alternating between shared tables. During insertion, the primary hash function is used first to determine a slot in the first table. If a collision is encountered, the secondary hash function is used to insert into the second table instead. This process continues with subsequent hash functions targeting additional tables in case of persisting collisions. Due to the relatively low collision rates encountered for big data workloads, only a small number of tables are required.

Listing \ref{list:software-memory-disambiguation} demonstrates an example code, which is based on the proposed coroutine-based framework (details in Section \ref{sec:programming_paradigm}) for simplification. Before initiating asynchronous memory requests that could violate ordering, the program calls the corresponding function to check for conflicts with outstanding requests (line 30). The function queries the target address in the hash table (line 7). If no entry exists, the to-be-accessed address is inserted into the hash table (line 13). Otherwise, the current coroutine is suspended and its handle is appended to the hash table entry (line 8-11). When other requests to the same address are completed, the suspended coroutine is awakened to resume (line19-21). In the future, advancements in compiler technology will potentially automate the insertion of such conflict-handling code, further reducing the burden on programmers. The hash table is located at the local DRAM. In fact most of it will be held in cache since the table is small and accessed frequently for memory extensive applications.  

\begin{lstlisting}[language=c, numbers=left, frame=single,
    xleftmargin=3em, framexleftmargin=2.5em, multicols=2, float,floatplacement=TB,
    caption=Pseudo Code of Software-based Memory Disambiguation,label={list:software-memory-disambiguation}]
// The provided code snippet is intended for
// illustrative purposes only. In an actual
// implementation, most operations would be
// encapsulated into awaitable objects and
// be highly optimized.
hash_map<uintptr_t, queue<coroutine_handle_t>> local_hashtable;
coro::task<> start_access(uintptr_t addr) {
  auto iter = local_hashtable.find(addr);
  if (iter != local_hashtable.end()) {
    // assume that `coroutine_handle` is 
    // the handle of current coroutine
    iter->second.push_back(coroutine_handle);
    suspend_current_coroutine();
  } else {
    local_hastable.insert(addr, queue<coroutine_handle_t>());
  }
}
void end_access(uintptr_t addr) {
  if (!local_hashtable[addr].empty()) {
    auto handle=local_hashtable[addr].front();
    local_hashtable[addr].pop();
    resume_the_coroutine(handle);
  } else {
    local_hashtable.erase(addr);
  }
}
void one_coroutine() { // usage example
  co_await start_access(mem_addr);
  co_await aload(spm_addr, mem_addr);
  // ...
  co_await astore(spm_addr, mem_addr);
  end_access(mem_addr);
  // ...
}
\end{lstlisting}

\subsection{Programming Framework\label{sec:programming_paradigm}}
While enabling much higher MLP and performance, AMI introduces two challenges about programming complexity. First, the asynchronous semantics are inherently less intuitive for programmers compared to synchronous programming. Programmers must manually check for completion of memory operations and handle callbacks. This event-driven style adds significant complexity. Second, rewriting existing code to extract sufficient MLP is difficult. Identifying independent memory operations and scheduling/synchronizing between them requires expertise. Overcoming these difficulties is essential to achieve the potential speedup of AMU. Thus, we propose a set of coding approaches as well as preliminary work on the compiler.

First, we exploit the async/await mechanism (i.e. coroutine) introduced in C++20 to preserve a synchronous-style programming model. By wrapping \textit{aload}/\textit{astore} in awaitable objects, our framework allows programmers to simply \textbf{co\_await} the corresponding awaitable objects when asynchronous memory access is needed, as shown in Listing \ref{list:amae-basic} line 48 and 50. Programmers no longer need to manually handle the asynchronous semantics introduced by asynchronous memory loads and stores.  This significantly reduces programming effort compared to directly using the AMI without higher-level abstraction. As shown in Listing \ref{list:amae-basic} line 44-57, the code that employs the proposed framework is significantly simpler than manually using AMIs (\ref{list:amae-basic} line 12-40), similar to the code in the "multi-thread version". Developers can continue writing code in an intuitive synchronous style while the underlying runtime transparently manages the asynchronous memory operations. While the programming framework is currently implemented in C++, porting it to other languages is feasible. Many languages nowadays such as Rust and Go also support coroutines.

Second, we argue that the common paradigm to exploit AMU is converting available loop-level parallelism (LLP) or request-level parallelism (RLP) into MLP. The opportunity of LLP generally appears in loops that have loop-independence. On the other hand, RLP is usually present in Internet servers that receive a large number of concurrent user requests. For programs that exhibit LLP or RLP, synchronous \textit{load/store} to far memory can be replaced with AMIs. Then, the threads can be interleaved effectively to hide the access latency. Lightweight coroutines executed with our framework incur lower overhead than OS threads for this purpose.

Listing \ref{list:amae-basic} presents a demonstration of how to convert a serial code to an AMI version. Initially, the sequential code (lines 1-3) is transformed into a multi-thread version (lines 4-11) by exploiting LLP/RLP. Subsequently, we generate an equivalent coroutine code (lines 12-40). The \textit{load/store} instructions to far memory are replaced with \textit{aload}/\textit{astore} and \textit{getfin}. The transformation between the multi-thread version and the coroutine-based code can be simplified by using the proposed framework (lines 43-54).

\begin{lstlisting}[language=c, numbers=left, frame=single,
    xleftmargin=2.5em, framexleftmargin=2.5em, multicols=2, float,floatplacement=TB,
    caption=Exploiting MLP with Asynchronous  ISA,label={list:amae-basic}]
// ----  original version  ----
for (i = 0; i < n; ++i)
    L[i] ^= i;
// ---- multi-thread version ----
void work_thread(int L[], int begin, int end) {
  for (i = begin; i < end; ++i)
    L[i] ^= i;
}
for (i = 0; i < n; i += slice)
  start_thread(work_thread, L, i, i + slice);
wait_all_thread_finish();
// ---- asynchronous memory access version ----
int *spm = (int*) spm_data_area;
int spm_free = 0;
auto make_task=[&](int begin, int end)->task<>{
  // alloc on spm
  int alloced = spm_free++;
  for  (i = begin; i < end; ++i) {
    int handle=aload(&spm[alloced],&L[i]);
    // this task will suspend until 
    // the memory access request complete.
    reschedule_to_other_task(handle);
    spm_scratch[alloced] ^= i;
    handle = astore(&spm[alloced], &L[i]);
    reschedule_to_other_task(handle);
  }
};
cfgwr(granularity, sizeof(int));
while (!is_all_task_finished()) {
  int handle = getfin();
  if (handle) { 
    // When the memory request finished, 
    // the corresponding task is resumed.
    resume(handle);
  } else {
    // There are no memory requests completed.
    // Thus, try to spawn a new task
    if (i < n) {
      spawn(make_task(i, i + slice));
      i += slice;
    }
  }
}
// ---- coroutine framework version ----
auto make_task=[&](int begin, int end)->task<>{
  int alloced = spm_scratch_free++;
  for  (i = begin; i < end; ++i) {
    co_await aload(&spm[alloced],&L[i]);
    spm_scratch[alloced] ^= i;
    co_await astore(&spm[alloced], &L[i]);
  }
};
vector<task<>> tasks;
for (i = 0; i < n; i += slice) {
  tasks.push_back(make_task(i, i + slice));
}
schedule_and_wait_all_task(tasks);
\end{lstlisting}

To further reduce porting overhead, we are developing an LLVM Pass that can automatically leverage aload/astore instructions to accelerate memory accesses for specified data-independent loops. The pass operates based on annotations provided by programmers in their source code. Specifically, programmers mark data as far memory and loops as data-independent using annotations. The pass then identifies these annotations and applies the approach of converting LLP to MLP described above to boost performance.

\subsection{System-level Issues}

\subsubsection{Context Switch\label{subsec:context_switch}}

The design of AMU assumes that user applications maintain exclusive access to AMU, as the process switching overhead is significant. In a multi-process environment, the operating system (OS) manages SPM similarly to vector registers. Thus, context switching requires saving the full SPM as well as several micro-architecture states. 
Additionally, when AMU-based programs and conventional programs are executed together on the same core, cache contention can occur. Thus, to ensure performance, AMU should be used exclusively as much as possible, and the OS should avoid co-scheduling cache-hungry programs with AMU-based programs on the same core while ensuring CPU affinity for AMU-based programs.


\subsubsection{Working with Synchronous Memory Access Code \label{subsec:work_with_legacy_code}}
Cooperating with existing synchronous memory access codes is another important consideration for AMU. Large applications typically contain code segments with different memory access characteristics, and legacy code that cannot be easily modified. While load and store instructions remain suitable for code with good data locality, other memory-intensive codes could benefit significantly from offloading memory operations to an AMU due to their poor data locality.

To facilitate cooperative execution, the programmer or compiler must analyze the phases and data structures of programs to determine the division of labor between synchronous and asynchronous approaches. Some programs can exhibit different access patterns in different phases - one with good locality and another with poor locality. This requires dynamically switching between synchronous and asynchronous memory access, with necessary cache flush instructions to maintain data consistency. Additionally, while some programs may not have clear-cut phases, the memory access patterns of different data structures within can vary greatly. Structures with good locality are better suited to use load/store and caches, while some structures could benefit from asynchronous memory accessing via AMU. This requires allocating these data structures into different memory regions.

\section{Evaluation\label{sec:evaluation}}

\subsection{Evaluation Methodology\label{subsec:methodology}}
\begin{figure}[tb]
  \setlength{\abovecaptionskip}{0pt} 
  \setlength{\belowcaptionskip}{-1pt}
\includegraphics[width=0.65\columnwidth]{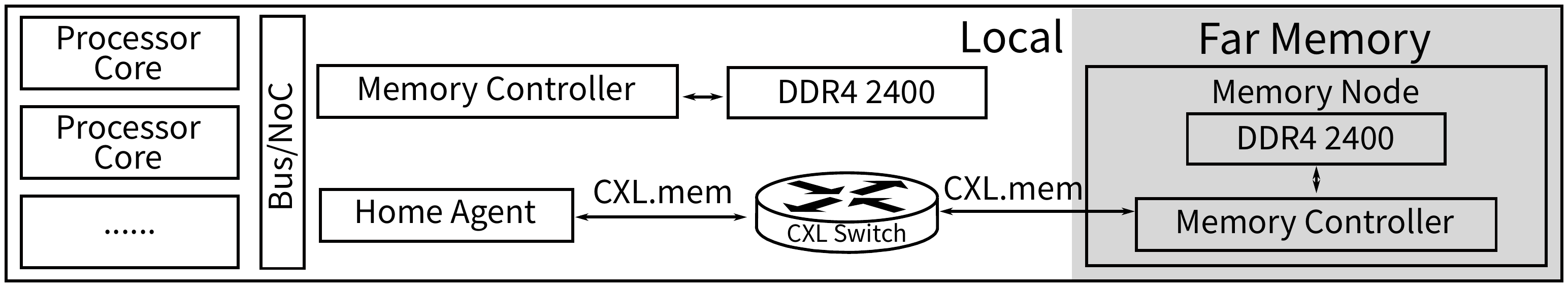}
\centering
\caption{The simulated systems with far memory\label{fig:simulated-topology}}
\end{figure}

\vspace{-5pt}
The proposed AMU is implemented by modifying the RISC-V OoO CPU model within the GEM5 simulator\cite{lowe2020gem5}. The simulated system architecture, as depicted in Figure \ref{fig:simulated-topology}, includes a remote memory node connected to the CPU via CXL as a far memory. 
Since this work is only concerned with memory access latency, CXL is modeled using the serial link model of gem5. It models the packet delay, which is dependent on the size of the packet, as well as the bandwidth limitation of the interconnect.
Internal details of CXL such as coherence protocols are not simulated. We integrate McPAT \cite{liMcPATIntegratedPower2009} to estimate power consumption. Additionally, an RTL model was developed using Chisel HDL to evaluate the logic resources overhead.

\begin{table}[tb]
  \setlength{\abovecaptionskip}{0pt} 
  \setlength{\belowcaptionskip}{0pt}
  \centering
  \footnotesize
  \begin{tabular}{lp{0.9\textwidth}}
    \toprule
    \multicolumn{2}{c}{\textbf{Configuration}}\\
    \midrule
    Processors & 3 GHz, 6-wide OoO, 512-entry ROB, 512 physical registers, 192-entry LSQ \\ 
    Cache & L1D/L1I: private, 32 KB, 16-way, 48-entry MSHR, delay = 4 cycles. 
    L2: private, 256 KB, 8-way, 48-entry MSHR, delay = 10 cycles \\
    Memory & DDR4\_2400\_8x8, 2 rank, 8 bank group, 32 bank \\
    \bottomrule
  \end{tabular}
  \caption{Gem5 Configuration}
  \label{table:gem5}
\end{table}

Evaluation is conducted under four configurations. The \textbf{Baseline} configuration, mimicking an Intel Golden Cove processor, is shown in Table \ref{table:gem5}. Moreover, we set up a configuration called \textbf{CXL Ideal (with BOP)} which is an "ideal" configuration with an L2 best-offset hardware prefetcher \cite{michaudBestoffsetHardwarePrefetching2016a}. In this configuration, the maximum number of in-flight memory requests is significantly increased by setting the entries of Miss Status Holding Register (MSHR) to 256 at each cache level. This configuration provides a useful target for comparison, approximating prior works that boosted hardware resources or pure hardware designs to increase out-of-order execution potential. Although impractical by current technology, it establishes an upper bound on attainable performance through conventional means. The \textbf{AMU} configuration refers to the proposed AMU architecture. In the evaluations, the size of the SPM is fixed at 64KB. The \textbf{AMU (DMA-mode)} is a limited \textbf{AMU} configuration, simulating the performance of external memory access engines. This configuration limits the number of IDs that the list vector registers can buffer to 1, forbidding the micro-ops to be executed speculatively.

Due to simulation time constraints, some simplifications are made. However, this does not change the general conclusion of our paper. First, cache and dataset sizes are decreased to ensure a reasonable data loading time. Despite reduced cache capacity, it still reflects the fact that applications enjoying far memory tend to have a significantly large working set, making it challenging for the cache to cover completely. Second, the evaluation uses single-core configurations, focusing on the ability of AMU to enable high outstanding memory requests. 

\subsection{Benchmarks\label{subsec:benchmarks}}
Multiple memory-bound benchmarks from various suites were chosen. The selected benchmarks include random access (GUPS), STREAM, binary search (BS), hash join (HJ) \cite{balkesen2013main}, hash tables (HT) \cite{David2015Asynchronized}, link list (LL) \cite{herlihy2012art} and skip lists (SL) \cite{David2015Asynchronized}, BFS, Integer Sort (IS), Redis and HPCG. 

The benchmarks were modified with AMI to exploit the MLP, following the programming paradigm described in Section \ref{sec:programming_paradigm}. Besides Redis and GUPS which are handwritten, other benchmarks are ported by using the proposed coroutine framework to reduce programming complexity. BS, HT, LL, SL, and Redis exploit request-level parallelism by launching multiple coroutines (256 for most, 128 for SL) to execute an independent key/value lookup task. Each coroutine sequentially generates random keys. The data structures being looked up are allocated in far memory. When accessing data structures in far memory, these workloads use AMI to initiate asynchronous memory access requests. The coroutines are interleave executed by the framework to hide far memory latencies. On the other hand, GUPS, HJ, HPCG, IS, and STREAM exploit loop-level parallelism, as they contain iterations that are independent of each other. Each coroutine executes a portion of the iterations. Most benchmarks have small (less than 64B) access granularity, except STREAM, IS, and HPCG which were evaluated for the benefits of large granularity. As these three benchmarks involve accessing contiguous memory, performance is improved by loading 512B or more of contiguous data into the SPM with each \textit{aload}/\textit{astore}.


\begin{table}[tb]
  \setlength{\abovecaptionskip}{0pt} 
  \setlength{\belowcaptionskip}{-1pt}
  \centering
    \footnotesize
  \begin{tabular}{p{0.2\columnwidth}p{0.8\columnwidth}}
    \toprule 
    \textbf{Name} & \textbf{Descriptions}\\
    \midrule
    Breadth First Search (BFS) & This implementation is from Graph500. The workload contains 16384 vertexes and 262144 edges. \\
    Binary Search (BS) & This program launches 256 coroutines to search random keys from a shared array. Each element is a 16B integer.  \\
    Random Access (GUPS) & Single node version of HPCC Random Access. The table been updated is located in far memory. \\
    Hashjoin (HJ) & This implementation is from \cite{balkesen2013main}. The hash table contains 16000 buckets.
    Each node of the list is 48 Byte. \\
    Hash Table (HT) & The chained hash table implementation is from ASYCLIB \cite{David2015Asynchronized}. Each node is same as the node of LL. \\
    HPCG & The OpenMP implementation of HPCG was selected. The matrices are allocated in far memory. \\
    Integer Sort (IS) & Integer Sort is a memory-bound random access benchmark from NAS Parallel Benchmark \cite{bailey1991parallel}. \\
    Hand-over-hand Linked List (LL) & This program lookup a number of keys from a hand-over-hand linked list \cite{herlihy2012art}. Each node contains 8 Byte key, 8 Byte value and a pointer to the next node. \\
    Redis & YCSB is used to benchmark the modified Redis. The Redis' single thread execution model is modified to support servicing concurrent requests. The buckets of the chained hash table are allocated in local memory, while the collision lists are allocated in far memory. \\
    Skip-list Lookup (SL) & The concurrent skip-list implementation is from ASYCLIB \cite{David2015Asynchronized}. Each node contains 32 Byte payload (including key, value and other metadata required by Skip-list) and 15 pointers. \\
    STREAM & The major arrays are allocated in far memory. \\
    \bottomrule
  \end{tabular}
  \caption{Description of benchmarks}
  \label{table:footprint}
\end{table}

\subsection{Performance and Power Evaluation\label{subsec:performance}}
\vspace{-5pt}
\begin{figure*}[tb]
\setlength{\abovecaptionskip}{0pt} 
\setlength{\belowcaptionskip}{-1pt}
\includegraphics[width=0.95\textwidth]{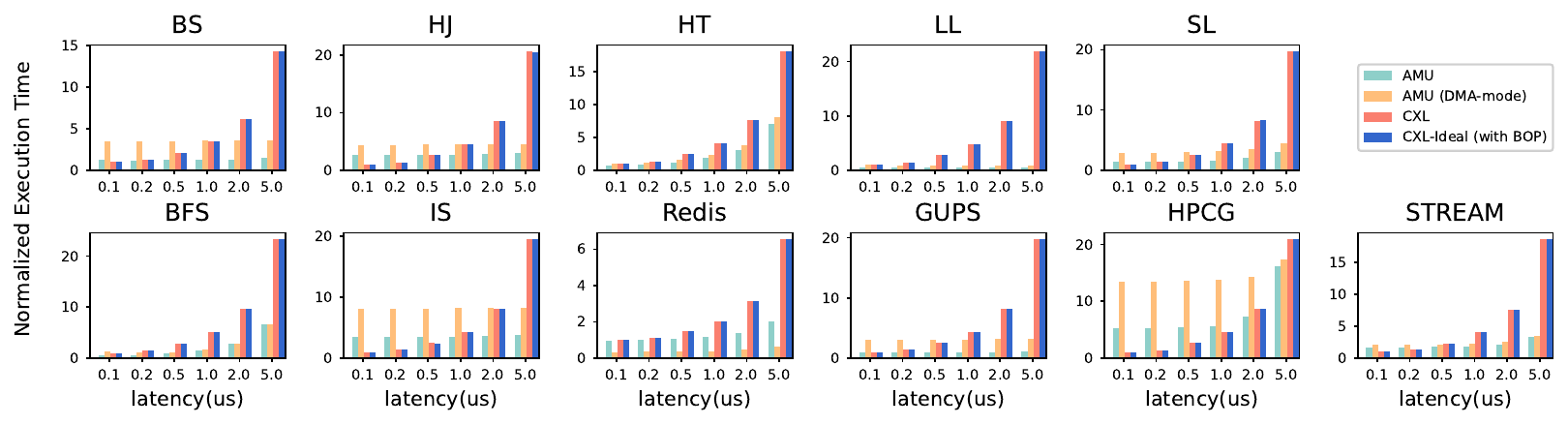}
\centering
\caption{Normalized execution times of benchmarks.\label{fig:normed_interval_per_bench} The x-axis represents the access latency introduced by far memory. The y-axis shows the normalized execution time. The execution time is normalized to the baseline configuration under 0.1$\mu$s far memory latency. 
}
\end{figure*}

\begin{figure*}[tb]%
  \centering
  \includegraphics[width=0.95\textwidth]{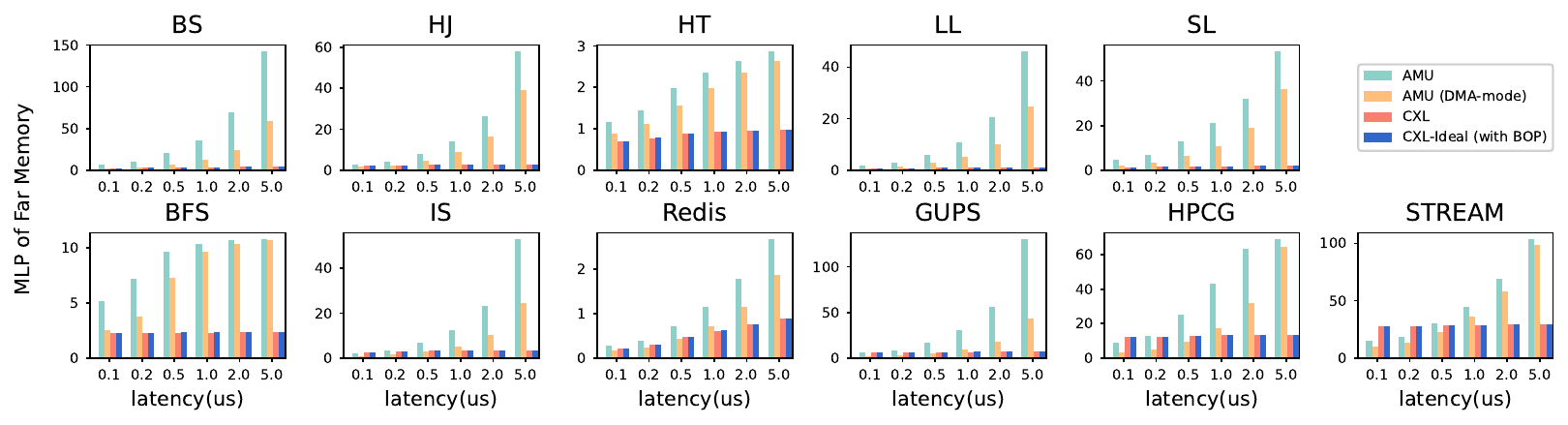}
  \caption{MLP of benchmarks. The x-axis represents the access latency introduced by far memory. The y-axis shows the MLP.}
  \label{fig:mlp_results}
\end{figure*}


Figure \ref{fig:normed_interval_per_bench} shows the normalized execution time of the benchmarks (the lower, the better). The far memory latency is adjusted to various values to simulate different far memory devices. For most cases, AMU maintains relatively constant performance as latency increases, demonstrating its ability to mitigate latency impacts. 

\textbf{AMU} exhibits performance advantages in most benchmarks when the additional latency caused by far memory exceeds 500 ns. For BS, BFS, GUPS, HT, and LL, \textbf{AMU} performs well when additional latency is only 0.2 $\mu$s. These workloads primarily involve random memory access, preventing them from benefitting from caches. Their performance is limited by the random access capability of hardware. Additionally, irregular access patterns diminish hardware prefetcher effectiveness, hurting performance. As AMU enables issuing huge numbers of outstanding requests, it alleviates these bottlenecks of workloads. For IS, which accesses memory sequentially more often, 
AMU only outperforms other configurations when additional latency is over 1 $\mu$s. 

The \textbf{AMU} demonstrates significant advantages over traditional external memory access engines simulated as \textbf{AMU (DMA-mode)}. As engines located outside the core cannot leverage out-of-order execution and have higher latency and overhead, AMU avoids these disadvantages through specialized microarchitectural design within the processor. By supporting out-of-order execution and the proposed batching mechanisms within the core, the AMU is able to significantly reduce overhead, making fine-grained asynchronous memory access practical.

Figure \ref{fig:mlp_results} shows the average number of in-flight memory requests (MLP), which directly demonstrates a key benefit of the AMU. As latency increases, AMU-based workloads exhibit a corresponding rise in average MLP, as asynchronous memory access instructions enable applications to schedule more overlapping independent instruction streams for execution. With higher latency, more coroutines can be interleaved due to the ability to asynchronously initiate additional memory requests. The improvement in MLP effectively mitigates the impact of increased latency. In contrast, the performance of original codes relies heavily on short latency due to the limitations of hardware resources and synchronous semantics. Their MLP lacks scalability and does not improve as far memory latency rises.

\begin{figure*}[tb]%
  \centering
  \includegraphics[width=0.95\textwidth]{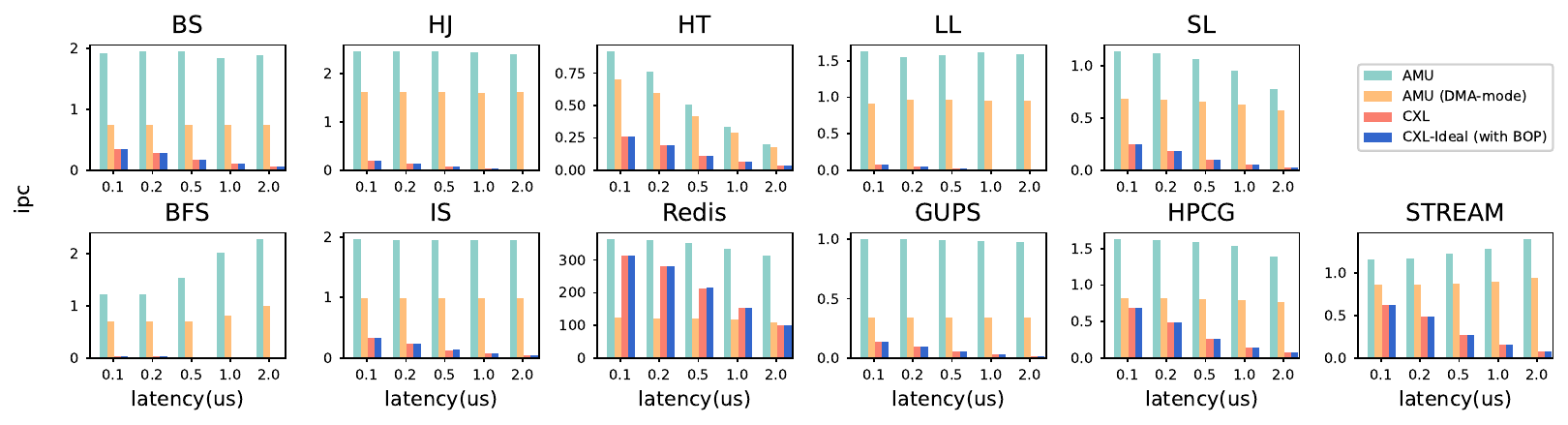}
  \caption{IPC of benchmarks. The x-axis represents the access latency introduced by far memory. The y-axis shows the IPC.}
  \label{fig:ipc_results}
\end{figure*}

Figure \ref{fig:ipc_results} shows the IPC of the benchmarks. It can be seen that adopting AMI significantly improves IPC. This demonstrates that the AMI, unlike traditional load/store instructions, does not stall for a long time in the ROB. Instead, they are committed very rapidly. This confirms that the proposed design is effective at reducing the consumption of critical hardware resources.

\begin{figure*}[tb]%
\centering
\includegraphics[width=0.95\textwidth]{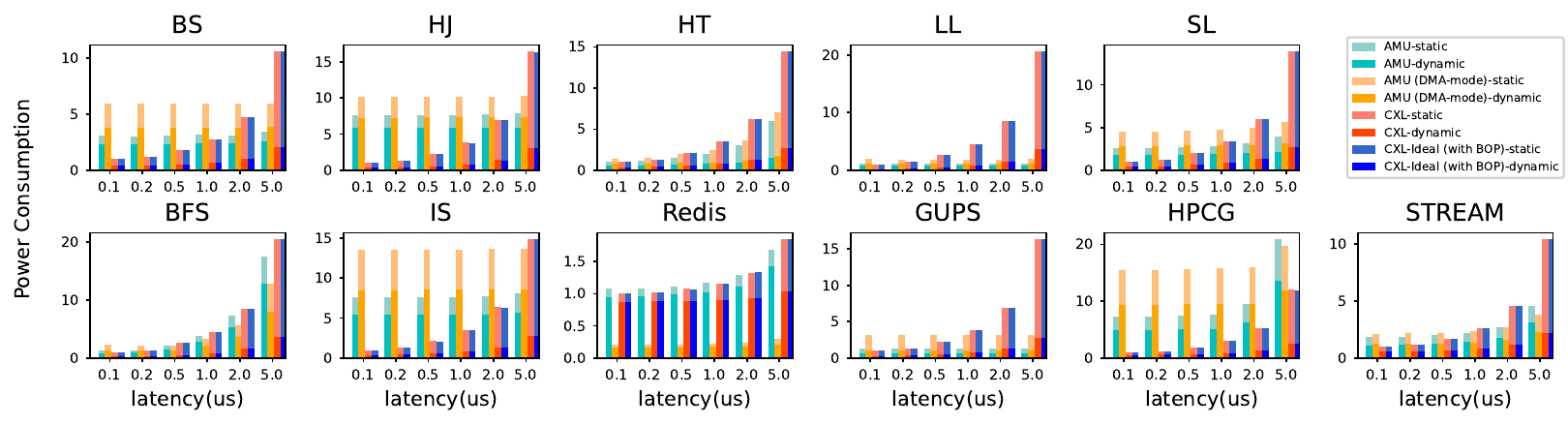}
\caption{Normalized power consumption of benchmarks. The y-axis shows the power consumption which is normalized to the baseline configuration under 0.1 $\mu$s far memory latency. The top (brighter) part represents static energy, while the bottom part represents dynamic energy}
\label{fig:power_consumption}
\end{figure*}

Figure \ref{fig:power_consumption} shows a breakdown of the power consumption. The additional power consumption of AMU is primarily due to the maintenance of metadata in the SPM and the increased instruction execution overhead caused by software-based instruction scheduling. Meanwhile, the performance enhancement offered by the AMU reduces the overall execution time, leading to a decrease in static power consumption and the counts of accessing power-consuming hardware resources such as ROB/IQ. This results in significant power consumption benefits. When the far memory latency is 500ns, the geometric mean of the power consumption of AMU relative to the baseline is 1.3, indicating that the power consumption benefits brought by performance improvement do not entirely cover this additional overhead. However, when the latency reaches 1 $\mu$s, the geometric mean of power consumption of AMU compared to the baseline reduces to 0.9, and the extra power consumption overhead is effectively compensated by the power consumption benefits achieved through performance improvement.

\begin{table}[tb]
\footnotesize
\centering
\begin{tabular}{c|llllllllllllll}
\toprule
\multirow{3}{*}{Latency ($\mu$s)} & \multicolumn{5}{c}{GUPS} & \multicolumn{5}{|c|}{HJ} & \multicolumn{3}{c}{STREAM} \\ \cline{2-14}
& \multicolumn{1}{c}{\multirow{2}{*}{CXL}} & \multicolumn{2}{c}{PF} & \multicolumn{1}{c}{\multirow{2}{*}{AMU}} & \multicolumn{1}{c}{\multirow{2}{*}{LLVM}} & \multicolumn{1}{c}{\multirow{2}{*}{CXL}} & \multicolumn{2}{c}{PF} & \multicolumn{1}{c}{\multirow{2}{*}{AMU}} & \multicolumn{1}{c}{\multirow{2}{*}{LLVM}} & \multicolumn{1}{c}{\multirow{2}{*}{CXL}} & \multicolumn{1}{c}{\multirow{2}{*}{AMU}} & \multicolumn{1}{c}{\multirow{2}{*}{LLVM}} \\
& \multicolumn{1}{c}{}                     & best       & config    & \multicolumn{1}{c}{}                     & \multicolumn{1}{c}{}                      & \multicolumn{1}{c}{}                     & best      & config     & \multicolumn{1}{c}{}                     & \multicolumn{1}{c}{}                      & \multicolumn{1}{c}{}                     & \multicolumn{1}{c}{}                     & \multicolumn{1}{c}{}                      \\
\midrule
0.1 & 1.00                                     & 4.52       & 2-128     & 0.96                                     & 0.69                                      & 1.00                                     & 0.86      & 128-128    & 2.69                                     & 1.84                                      & 1.00                                     & 1.64                                     & 13.02                                     \\
0.2 & 1.38                                     & 6.39       & 32-128    & 0.96                                     & 0.69                                      & 1.41                                     & 1.19      & 128-128    & 2.67                                     & 1.84                                      & 1.28                                     & 1.67                                     & 15.28                                     \\
0.5 & 2.54                                     & 12.98      & 8-0       & 0.97                                     & 0.70                                      & 2.61                                     & 2.21      & 128-32     & 2.68                                     & 1.85                                      & 2.28                                     & 1.74                                     & 22.44                                     \\
1.0 & 4.40                                     & 23.01      & 32-64     & 0.98                                     & 0.69                                      & 4.59                                     & 3.86      & 128-128    & 2.71                                     & 1.87                                      & 4.00                                     & 1.87                                     & 35.08                                     \\
2.0 & 8.21                                     & 43.73      & 16-4      & 1.00                                     & 0.70                                      & 8.61                                     & 7.24      & 128-32     & 2.79                                     & 2.13                                      & 7.63                                     & 2.18                                     & 64.14                                     \\
5.0 & 19.83                                    & 106.77     & 16-4      & 1.03                                     & 0.71                                      & 20.70                                    & 17.44     & 128-32     & 3.08                                     & 2.13                                      & 18.66                                    & 3.33                                     & 149.83                                    \\ \bottomrule
\end{tabular}
\caption{Normalized execution time of baseline (\textbf{CXL}), compiler-based software prefetch (\textbf{PF}), AMU, and LLVM-based AMU (\textbf{AMU}). The config list specifies the prefetch aggressiveness (x-y), where x indicates batching the memory accesses of x iterations and y indicates the depth of prefetching indirect memory access\label{tab:llvm}}
\end{table}

Table \ref{tab:llvm} compares the performance of a software prefetching scheme versus AMU. The software prefetching scheme utilizes a compiler-based software prefetching \cite{tranClairvoyanceLookaheadCompiletime2017}. The data shows that software prefetching requires careful tuning of prefetch aggressiveness due to the lack of feedback from hardware on the availability of prefetched data. This makes it difficult for software prefetching to adapt effectively to scenarios with highly variable memory access latencies like far memory.

Additionally, table \ref{tab:llvm} a preliminary evaluation of the AMU LLVM pass. The evaluation indicates the compiler-directed optimizations outperformed manually ported versions significantly (e.g. GUPS and HJ). 
On the other hand, for the compiler-based STREAM, performance was notably lower compared to the hand-optimized version using large-granularity asynchronous memory accesses. This is because the current compiler only supports 8B granularity asynchronous memory accesses.
This demonstrates that with continued advancement in compiler techniques, AMU still has the potential for higher performance gains.

Table \ref{tab:software-disambiguation} gives the overhead of memory disambiguation for two typical workloads. For HJ, the cost of memory disambiguation remains fairly stable around 5\%. For HT, the percentage of time spent on memory disambiguation is higher when the remote latency is low. However, as latency increases, this portion rapidly declines. Additionally, while there is a noticeable cost for memory disambiguation, the benefits from asynchronous memory access outweigh these overheads. Therefore, this overhead is acceptable. In future work, hardware-assisted techniques \cite{zhuSynchronizationStateBuffer2007} could be explored to further reduce the cost and unlock more of the potential of AMU.

\begin{table}[tb]
\centering
\footnotesize
\begin{tabular}{lllllll}
\toprule
Latency ($\mu$s) & 0.1     & 0.2     & 0.5     & 1.0     & 2.0    & 5.0    \\
\midrule
HJ        & 5.06\%  & 5.04\%  & 5.07\%  & 5.07\%  & 5.00\% & 4.95\% \\
HT        & 32.47\% & 29.04\% & 20.17\% & 13.89\% & 9.14\% & 3.95\% \\
\bottomrule
\end{tabular}
\caption{Execution time spent on software-based memory disambiguation \label{tab:software-disambiguation}}
\end{table}

\subsection{Hardware Overhead}
\vspace{-5pt}
The on-chip storage overhead introduced by AMU is relatively low, as AMU reuses existing hardware resources. First, the list vector registers can employ the existing register renaming mechanism and use the general-purpose physical vector registers. Second, the metadata maintained by ASMC are stored in SPM, which is a part of the existing cache. Thus, no additional storage overhead is required for metadata. 

The additional storage overhead of AMU is mainly internal state registers and relatively short queues. First, each state machine of AMU requires a 32-entry pending queue and several internal state registers. Second, ASMC prepares two list-vector-register-length buffers as a cache of the corresponding list. Third, there are two Uncommitted ID Registers in the ALSU. Therefore, the total overhead is only approximately a few KB and does not vary when the required MLP increases.

We implemented the AMU on NanHu-G, which is the second-generation of the open-source high-performance OoO RISC-V processor XiangShan\cite{micro2022xiangshan}. NanHu-G is a 4-issue OoO core with speculation execution and 96 ROB entries. The AMU design was evaluated on an FPGA platform to analyze its hardware cost.  Furthermore, the area overhead of integrating the AMU is evaluated using Synopsys Design Compiler under the TSMC 28nm HPC+ process technology. Table \ref{tab:fpga} shows the resource utilization compared to NanHu-G.
The results indicate that the AMU can be efficiently implemented on modern processor cores with modest resource overheads.

\begin{table}[tb]
\centering
\footnotesize
\begin{tabular}{c|ccccc||c|c}
\toprule
\multirow{2}{*}{FPGA} & LUT (as Logic)   & LUT (as Memory)  &  FF  & BRAM   & URAM & \multirow{2}{*}{ASIC} & Area \\
\cline{2-6}\cline{8-8}
& +6.9\%     & +8.5\% & +4.5\% & +0\% & +0\% & & +6.67\% \\ 
\bottomrule
\end{tabular}
\caption{Resource utilization compared to NanHu-G \label{tab:fpga}}
\end{table}

\section{Related Work}
\subsection{Multi-threading/Many-core Architectures\label{subsec:many-core}}
Significant prior work attempted to provide massive hardware threads that can be used to hide memory access latency. For instance, Cray MTA\cite{mta2}/XMT\cite{ELDORADO} provides a large number of hardware threads. They offer hardware-based context-switch when the thread faces memory stalls.
Typical many-core processors such as Sunway \cite{fu2016sunway} and SmarCo \cite{fan2018smarco} also provide a massive number of hardware threads which is sufficient for hiding long memory access latency.
Intel PIUMA\cite{aananthakrishnanIntelProgrammableIntegrated2023} proposed the multi-threaded core which offers multiple hardware threads by round-robin executing them. However, supporting massive light-weight hardware threading comes at a higher cost, and mainstream general-purpose processors typically only support a small number of threads per core. To address this, Feliu et al proposed VMT\cite{feliuVMTVirtualizedMultiThreading2022}, which offers multiple virtual hardware threads by interleaving executing them on two hardware threads. However, the total number of the hardware threads remains limited. 
The key difference of AMU is that it aims to provide similar multithreaded high concurrency memory access capabilities as prior domain-specific designs within the general-purpose OoO core at a lower hardware and compatibility cost. AMU is an independent unit within the processor. It does not impact programs that do not use AMU, while enabling high MLP for programs that can utilize the provided mechanisms.

\subsection{Hybrid Memory/NUMA\label{subsec:hybrid_memory}}
Optimizing far memory accesses shares similarities with prior work on optimizing access in hybrid memory and non-uniform memory access (NUMA) systems. Hybrid memory systems commonly contain both faster DRAM and slower NVM. Likewise, NUMA architectures typically incur significant latency variations between local and remote node memory access. Many prior works explored memory allocation, data migration, and prefetching techniques to mitigate the impact of disparate access latencies in these systems. Memory allocation optimizations \cite{liu2020tc} aim to improve performance by placing frequently accessed objects in the faster memory. Data migration optimizations \cite{openmem, openmem2, Salkhordeh2016date, openmem2021} monitor access patterns and migrate hot data to the faster memory. Prefetching techniques \cite{Fedorov2017memsys, pan2021date, Haifeng2023HoPP} predict and prefetch data from the slower memory. These methods are orthogonal and complementary to AMU. They typically focus on page-granularity memory management and solve the problem from a more system-oriented perspective. The performance of synchronous load/store can be optimized by these techniques. In contrast, AMU focuses on providing tolerance for longer latencies through microarchitectural design. The major design goal of AMU is addressing the higher far memory access latency.

\section{Conclusion and future work\label{sec:conclution}}
\vspace{-5pt}

This paper introduces an ISA extension for asynchronous memory access along with the supporting hardware unit called AMU. Addressing the MLP limitation of modern OoO processors, the novel design adds support for full asynchronous mechanism with a small amount of hardware overhead. The design enables applications to exploit massive MLP to hide the latency of far memory. The evaluation reveals that memory-bound applications can exploit the full potential of far memory by utilizing the asynchronous programming paradigm.

This paper only provides fundamental instructions and structures to enable asynchronous memory access. Additional research are required in the areas of operating systems and compilers. In the future, more instructions can be added to further improve performance or provide more advanced memory access functions. For example, it is possible to add instructions to initiate a request with a group of memory operations together. As another example, instructions can be added to assist the software in scheduling the execution of other instructions after issuing an asynchronous access request (e.g. instructions to assist the scheduling of coroutines).

Furthermore, the proposed asynchronous design can be easily extended to support more functional memory extensions, including complex access patterns and Processing-In-Memory (PIM) mechanism. This extendibility is contingent upon the far memory subsystem supporting broader memory semantics. 

\bibliographystyle{ACM-Reference-Format}
\bibliography{refs}

\end{document}